\documentstyle[sprocl,epsf]{article}

\def\Journal#1#2#3#4{{#1} {\bf #2}, #3 (#4)}

\def\ZHTF{\em Pisma ZhETF}

\def\SNC{\em Suppl. Nuovo Cimento}

\def\PGNP{\em J.Phys. G.Nucl. Phys.}
\def\NPB{{\em Nucl. Phys.} B}
\def\PLB{{\em Phys. Lett.}  B}

\def\ZPC{{\em Z. Phys.} C}
\def\be{\begin{equation}}
\def\ee{\end{equation}}
\def\bea{\begin{eqnarray}}
\def\eea{\end{eqnarray}}
\begin{document}

\title{TRANSFORMATIONS OF THE HADRONIC AND SUBHADRONIC SUBSTANCES
UNDER EXTREME CONDITIONS}

\author{E.L. FEINBERG }

\address{P.N. Lebedev Physical Institute, 53 Leninsky Prospect, Moscow,
117924}

\maketitle\abstracts{
Very dense and/or hot hadronic substance (e.g. the one with energy density
greatly exceeding that of a normal nucleus) transforms itself into a
subhadronic substance which obeys macroscopic classical physics,
in particular suffers phase transitions. The most popular Single Phase
Transition Model (SPTM) assumes that the new phase is the
Quark Gluon Plasma ($QGP$) consisting of deconfined,
chiral symmetric, pointlike "current" quarks $q$ and gluons $p$
of Quantum Chromodynamics ($QCD$). This paper is devoted to another,
{\em Double} Phase Transition Model (DPTM) according to which
hadronic substance ($H$) and $QGP$ transform one into another
via an intermediate phase consisting of deconfined {\em constituent}
massive quarks $Q$ which for brevity sake we call also
equivalently {\em valons} ($Q$, valonic phase) with broken chiral symmetry
(plus pions as Goldstone particles). I. e. we consider the
phase transformation chain $H{\longleftrightarrow}
Q{\longleftrightarrow}{QGP}$ instead of usually assumed
$H{\longleftrightarrow}{QGP}$. The phase transition
$H{\longleftrightarrow}Q$ is the Hagedorn one and corresponds to the
Hagedorn temperature. Connection with the relativistic heavy ion collision is
discussed. $H{\longleftrightarrow}Q$ transformation may take place even
at low (e.g. Dubna) energies.}

\section{Introduction}
There still exist two prejudices.

1) Classical treatment of hadronic matter at extreme conditions
is considered a rough vulgarisation while it is actually fully proper.
 Two simplest arguments for this are:

\noindent {\em i)} Here in a single collision
very many particles appear, i.e. the number of excited degrees
of freedom becomes huge as well as the total quantum numbers
leading to classical physics.

\noindent {\em ii)} In strong interaction field theory
all elements of the Fock column are equally important.
Even the vacuum becomes filled up by spontaneously appearing
particles, its energy density $ \epsilon $ and pressure $p$
are estimated within $QCD$ as $- \epsilon_{vac}={p_{vac}} \simeq 0.5 - 1 \;
\mbox{GeV/fm}^3$, i.e.   average energy density of a
continuous medium of close packed nucleons, $\epsilon \approx \epsilon_N
\approx 0.5 \:\mbox{GeV/fm}^3$.  Thus in certain cases the ultramicroworld
may be described by macrophysics. One can even suggest that $any$
field theory at high $\epsilon$, if interaction becomes strong, should reduce to
the classical one.

2) After appearance of $QCD$ with its pointlike massless quarks
$q$ overwhelming majority of high energy physicists
seemingly forgot about massive constituent quarks
(or {\em valons}) $Q$ with which the quark idea had begun.
Systematics and many properties of baryons (three valons)
and mesons ($Q \overline {Q}$) were explained with masses
$\quad m_{Q(u)} \simeq m_{Q(d)} \simeq 310-340\:\mbox{MeV}$, \quad $m_{Q(s)}\approx
510\: \mbox {MeV}$ and average radius $r_Q\simeq 0.3\:\mbox{fm}$.  As a
meaningful exclusion one can mention Kennet Wilson Ohio University group who
recently reported on "the fruit of four years struggle and effort to build a
bridge between constituent quark model and $QCD$~\cite{lf}".

In the present paper, similarly to$\!\phantom{d}^{2-6}$,
$\; Q$ is treated as a real particle (imagined as $q$ dressed
by a dense cloud of $\; q \overline {q}\;$ pairs and gluons)
and phases of hadronic nature substance are considered classically.
%%%%%%%%%%%%%%%%%%%%%%%%%%%%%%%%%%%%%%%%%%%%%%%%%%%%%%%%%%%%%%%%%%
\section{Single Phase Transition Model (SPTM).}\label{sec:sptm}
%%%%%%%%%%%%%%%%%%%%%%%%%%%%%%%%%%%%%%%%%%%%%%%%%%%%%%%%%%%%%%%%%%
SPTM follows from the Hagedorn observation~\cite{hag} that the mass spectrum
of existing hadrons obeys $\rho(m) dm \sim exp( \frac{m}{T_H})dm$ (supported
later in various theoretical approaches).  According to latest fitting, $T_H
\approx 150\:\mbox{MeV}$~\cite{to}.  Since Boltzman average of any function
$f(m)$ is $\overline {f} \sim \int \! f(m) \rho(m) exp( \frac{- \epsilon
(m)}{T}) dm_q$ it diverges at $T \ge T_H$. Thus $T_H$ is the maximum
possible temperature for hadronic phase $H$.  Above it we must have another
phase practically unanimously believed to be $QGP$.  Thus $T_H$ should
coincide with temperatures of deconfinement, $\; T_d$, and chiral symmetry
restoration, $T_{ch}$:  $T_d=T_{ch} \equiv T_H$.
%%%%%%%%%%%%%%%%%%%%%%%%%%%%%%%%%%%%%%%%%%%%%%%%%%%%%%%%%%%%%%%%%%%
\section{Double Phase Transition Model (DPTM).}\label{sec:dptm}
%%%%%%%%%%%%%%%%%%%%%%%%%%%%%%%%%%%%%%%%%%%%%%%%%%%%%%%%%%%%%%%%%%%
However various arguments, e.g. existence of $Q$, point to existence
of additional mass scale and thus to possibility ot
$T_d \neq T_{ch}$~\cite{shur}.
Field theoretical analysis shows~\cite{pis,boch}
that if so then $T_{ch}\ge T_d$.
In fact, consider a simple pattern~\cite{fein}.
Let a nucleus with its $\epsilon_A=0.15\; \mbox{GeV/fm}$
to be compressed nearly thrice when its $\epsilon$
becomes equal to a nucleon one, $\epsilon_N=0.5\; \mbox{GeV/fm}$~\cite{soh},
i.e. until all nucleons are close packed.
Now each $Q$ confined within them can go over
to any adjacent nucleon, i.e. becomes {\em deconfined}
and they all form a gas of massive valons (plus pions as Goldstone particles).
 Further compression up to
%%$\epsilon= \epsilon_Q \sim \frac{m_Q}{\displaystyle \frac43  \pi r_Q^3} \sim\:
$\epsilon= \epsilon_Q \sim {m_Q}/\frac 43 \pi r_Q^3 \sim
3\:\mbox{GeV/fm}^3$ makes $Q$ close packed and enables $q$ quarks to go over
from one valon to any adjacent one thus deconfining them and forming $QGP$ .
This simplified scheme clarifies attempts$\!\phantom{d}^{2-6}$  to construct
a bag type model in which $Q$ phase was expected to be
intermediate.  However their results showed negligible role of $Q$ phase.

These works were reconsidered within the same
bag type thermodynamical model (calculation of partition function
for three phases with bag constants $B_Q$ and $B_q$ for $Q$
and $QGP$ phases). Herefrom for each $T$ and chemical potential $\mu$
the stable phase can be determined as the one with largest pressure.
However the physical approach to the choice of the free parameter $B_q$
was different. It was put equal to $p_{vac}$. This DPTM
has given expected result~\cite{hhm}: the phase transition
$H{\longleftrightarrow}QGP$ proceeds via the intermediate valon phase,
$H{\longleftrightarrow}Q{\longleftrightarrow}{QGP}$.  At some set of the
model parameters considered as "standard" ($B_Q= 50\:\mbox{MeV/fm}^3$,
$B_q=p_{vac}=0.5\:\mbox{GeV/fm}^3, \quad r_Q=0.3\:\mbox{fm}$) two phase
transitions for $\mu=0$ occure at $\quad T_d(\equiv T_H)=140\:\mbox{MeV}, \quad
T_{ch}=200\:\mbox{MeV}$ (Fig.1). The width of the $Q$ corridor at almost any
$\mu$ is few tens MeV.

Special efforts were directed to testing stability of this conclusion against
variation of free parameters. Increase of $B_Q$ narrows the corridor (Fig.2)
but leaves result qualitatively unchanged at least for $\mu \le 1\:
\mbox{GeV}$ (this can be compensated by assuming $p_{vac}\sim
1\:\mbox{GeV/fm}^3$: increase of $B_q$ widens the corridor). Variation of $H$
phase description within Hard Core Model (HCM) and Mean Field Approximation
(MFA) with two versions of nucleon interaction potential (Fig.3), account in
the $H$ phase of exciting 30 resonances (Fig.4) besides initially taken into
account merely $N, \Lambda, \pi$ and $K$, as well as variation of $r_Q$ do not
tell to any extent essentially.
%%%%%%%%%%%%%%%%%%%%%%%%%%%%%%%%%%%%%%%%%%%%%%%%%%%%%%%%%%%%%%%%%%%%%%%%
\section{Futher results}
%%%%%%%%%%%%%%%%%%%%%%%%%%%%%%%%%%%%%%%%%%%%%%%%%%%%%%%%%%%%%%%%%%%%%%%%
1) Duration of various phases in the process of expansion of initial
fireball generated by two nuclei collision as well as of mixed phases at
$T_d$ and $T_{ch}$ calculated according to the Bjorken simplified hydrodynamics
is presented in Fig.5. In DPTM it markedly differs from SPTM. Since $T_d$
is very close to freeze out temperature $T_f \approx 130\:\mbox{MeV}$ the pure
$H$ phase lasts very short time. A predominantly longer time is spent for
$Q$ and mixed $H+Q$ phases~\cite{bar}. This should tell quantitatively
on direct photon and dilepton and strangeness production.

2) As has been noticed in~\cite{leo}, the entropy per baryon $S/N_B$ is
not continuous at phase transition is SPTM. The same holds for DPTM.
This was cured in SPTM~\cite{leo} by making $B_q$ depending
on $T$ and $\mu$. In our treatment this is hardly proper since $B_q=p_{vac}$.
Accordingly in DPTM the same was done only for $B_Q$ by making it $T$ and $\mu$
dependent within $T_d \le T \le T_{ch}$.

%%%%%%%%%%%%%%%%%%%%%%%%%%%%%%%%%%%%%%%%%%%%%%%%%%%%%%%%%%%%%%%%%%%%%%%
%3) Inclusion of $\approx $ 30 hadronic resonances gave final particle
%content which
%differs weaker from the results for few hadron species taken into account
%initially than in SPTM.
%It deserves mentioning that
%%%%%%%%%%%%%%%%%%%%%%%
3)Relative content of strange
particles $\frac{K^-}{\pi^-}, \quad \frac{\Lambda^-}{\pi^-}, \quad
\frac{\overline {\Lambda} }{\pi^+}$ within DPTM is smaller than in SPTM
(due to lower temperature of hadronic phase appearence), save the special
case of $\frac{K^+}{\pi^+}$: here it is also smaller for
$\mu \le 200\:\mbox{MeV}$ but for $\mu>200\:\mbox{MeV}$ overcomes ($\pi $
producton in SPTM by hadronic resonance decay plays more important role than in
DPTM, again due to lower $T$).  
%%%%%%%%%%%%%%%%%%%%%%%%%%%%%%%%%%%%%%%
\section{Concluding remarks.}
The above said seemingly supports the idea of existence
of the valonic $Q$ phase and two phase transitions with
$T_d \neq T_{ch}$. The decisive condition for it is
sufficiently large ratio $\beta \equiv \frac{B_Q}{B_q} \ge 5$.
One of the most dubious elements of this
thermodynamical approach is the assumed independence of the
valon mass $m_Q$ of temperature. It seems plausible
that, like for other hadrons, it depends on quark condensate,
$m_Q \sim \langle 0| \overline {q}q|0\rangle ^{1/3}$ and diminishes
with $T$ increasing within $T_d<T<T_{ch}$.
Physically this can be imagined as undressing of $Q$,
loosing parts of its $q \overline {q} g$ cloud.
Estimates show that diminishing of $m_Q$ displaces $T_{ch}$ to higher values.
However on the other hand the assumed constancy of $B_Q$
within the same interval is also dubious.
Diminishing of $m_Q$ means its development in direction to $q$ mass,
and thus of $B_Q$ to $B_q$ which influences $T_{ch}$ in inverse sense. These
effects are now under investigation.

It deserves stressing that $\epsilon \sim 3 \epsilon_A$
necessary for coming to $Q$ phase
needs central collision of two identical nuclei
Lorentz contracted only $\sim\!1.5$ times
(even if we neglect additional contraction due to shock wave).
I.e. it can take place even at rather low energies,
$E_{Lab} \ge 4\:\mbox{GeV/nucleon}$.

It is to be added that recently O.K. Kalashnikov build
a field theoretical model starting from the complex Lagrangian
and imposing a special condition on coupling constant behaviour~\cite{kal}.
This quite a different approach also gave a wide
$Q$ phase corridor in $\mu-T$ plane (with smaller $Q$ masses).
%%%%%%%%%%%%%%%%%%%%%%%%%%%%%%%%%%%%%%%%%%%%%%%%%%%%
\section*{Acknowledgments}
%%%%%%%%%%%%%%%%%%%%%%%%%%%%%%%%%%%%%%%%%%%%%%%%%%%%%%%
I am grateful for collaboration in developing DPTM to O.D. Chernavskaya
and for various fruitfull discussions criticism to many colleagues,
especially to H. Satz and H. Leutwyler.
This paper was supported in part by the Russian Foundation for
Fundemental Reseaches, grants Nos 94-02-15558 and 94-02-3815.
%%%%%%%%%%%%%%%%%%%%%%%%%%%%%%%%%%%%%%%%%%%%%%%%%%%%%%%%%%%%%%%%

%%%%%%%%%%%%%%%%%%%%%%%%%%%%%%%%%%%%%%%%%%%%%%%%%%%%%%%%%%%%%
%\vspace*{5cm}
\newpage 
%%%%%%%%%%%%%%%%%%%%%%
   \begin{figure}
   \vspace{-5.5cm}
   \hspace*{4.cm}
   \epsfxsize=3.6truein
   \epsfysize=4.truein
   \epsffile{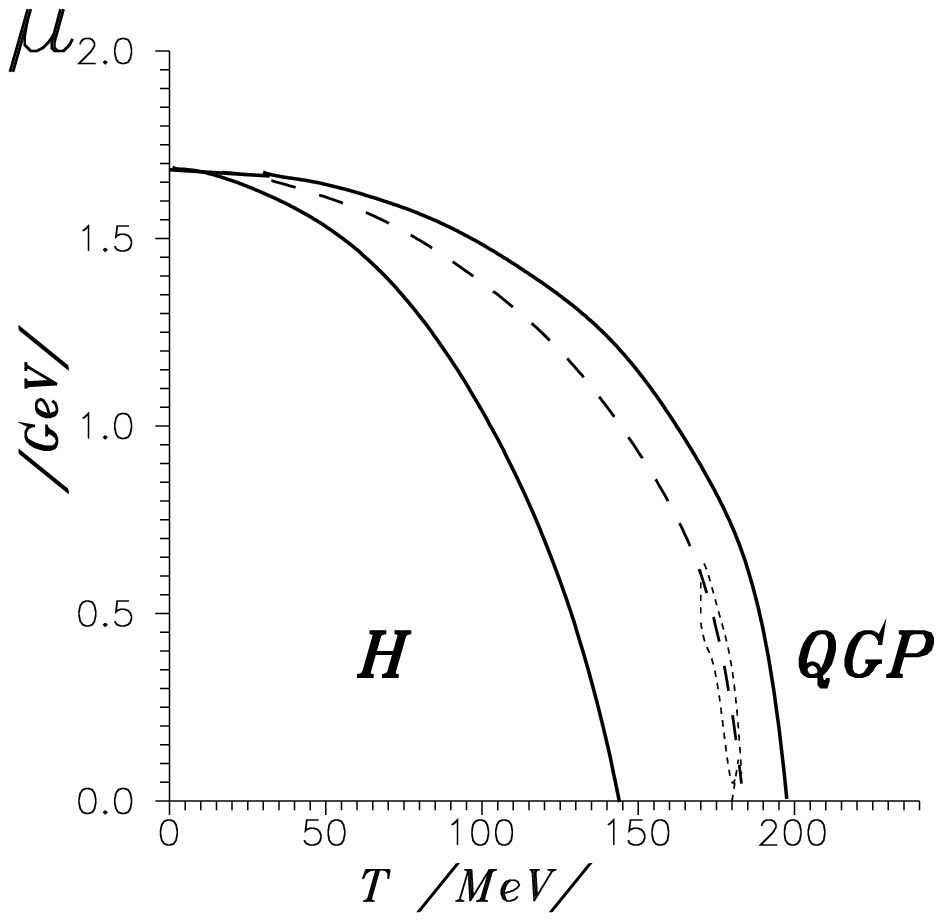}
~\\~\\
~\\
~\\
{\small Figure 1: The $\mu-T$ phase diagram according to SPTM (dashed)
and DPTM (solid lines) with $\beta $ =10. Dots bound the $Q$-phase for
$\beta \approx $ 3.} 
\end{figure}
%%%%%%%%%%%%%%%%%%%%%%%%%%%%%%%%%%%%%%%%%%%%%%%%%%%%%
\vspace*{-4.5cm}
%   \begin{*figure}
   \hspace*{3.1cm}
   \epsfxsize=3.6truein
   \epsfysize=4.truein
   \epsffile{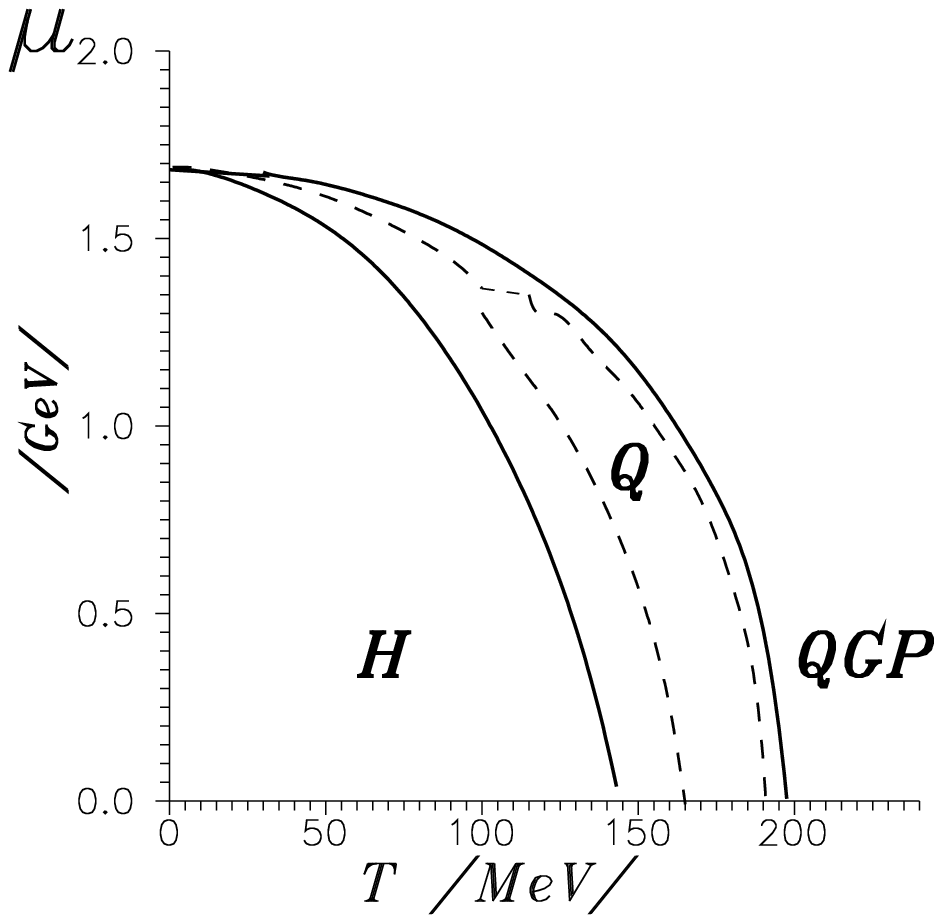}
~\\
~\\
~\\
\begin{center}
{\small Figure 2: DPTM transition curves from~\cite{hhm} with $B_Q=50$
MeV/fm$^3$ (solid) and $B_Q=100$ MeV/fm$^3$ (dashed).} 
\end{center}
%%%%%%%%%%%%%%%%%%%%%%%%%%%%%%%%%%%%%%%%%%%%%%%%%%%%%%
\newpage
%%%%%%%%%%%%%%%%%%%
\begin{figure}
\vspace*{-4.cm}
   \hspace*{4.cm}
   \epsfxsize=3.6truein
   \epsfysize=4.truein
   \epsffile{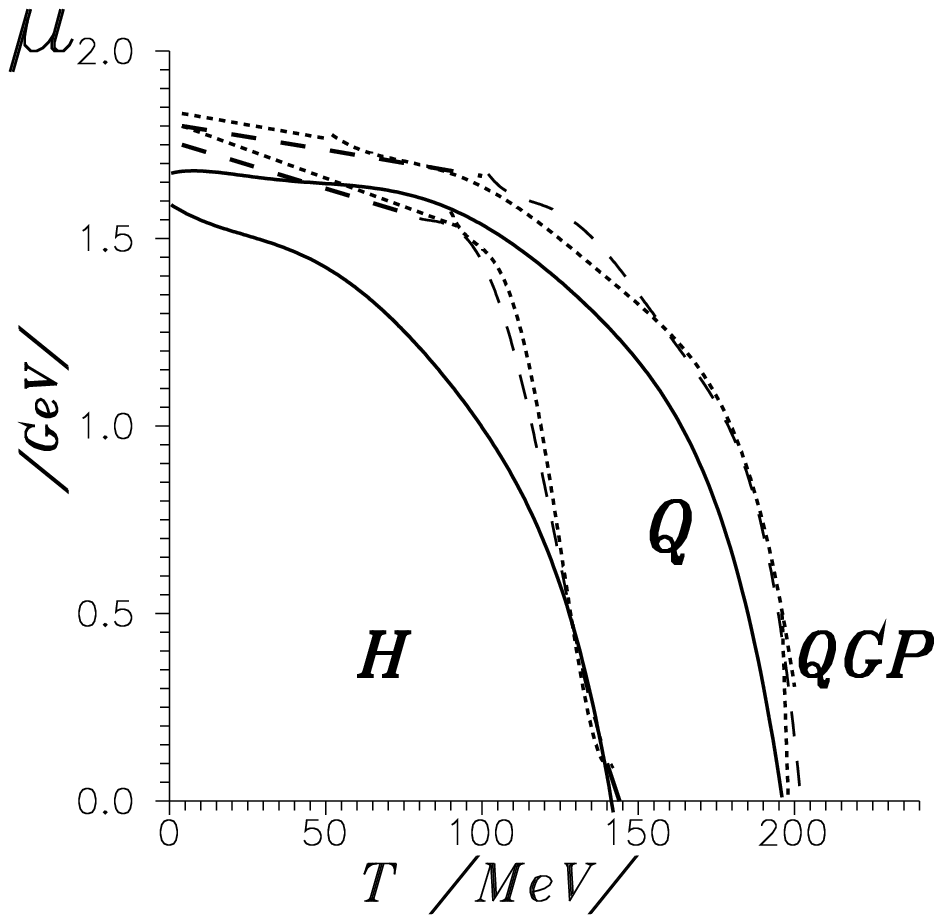}
~\\
~\\
~\\
\begin{center}
{\small Figure 3: DPTM transition curves for various nucleon
interaction description: HCM (solid) and two versions of MFA potential
for $H$-phase from ~\cite{bar} (dashed and dotted).}
\end{center}
\end{figure}
%%%%%%%%%%%%%%%%%%%%%%%
\vspace*{-5.cm}
   \hspace*{3.1cm}
   \epsfxsize=3.6truein
   \epsfysize=4.truein
   \epsffile{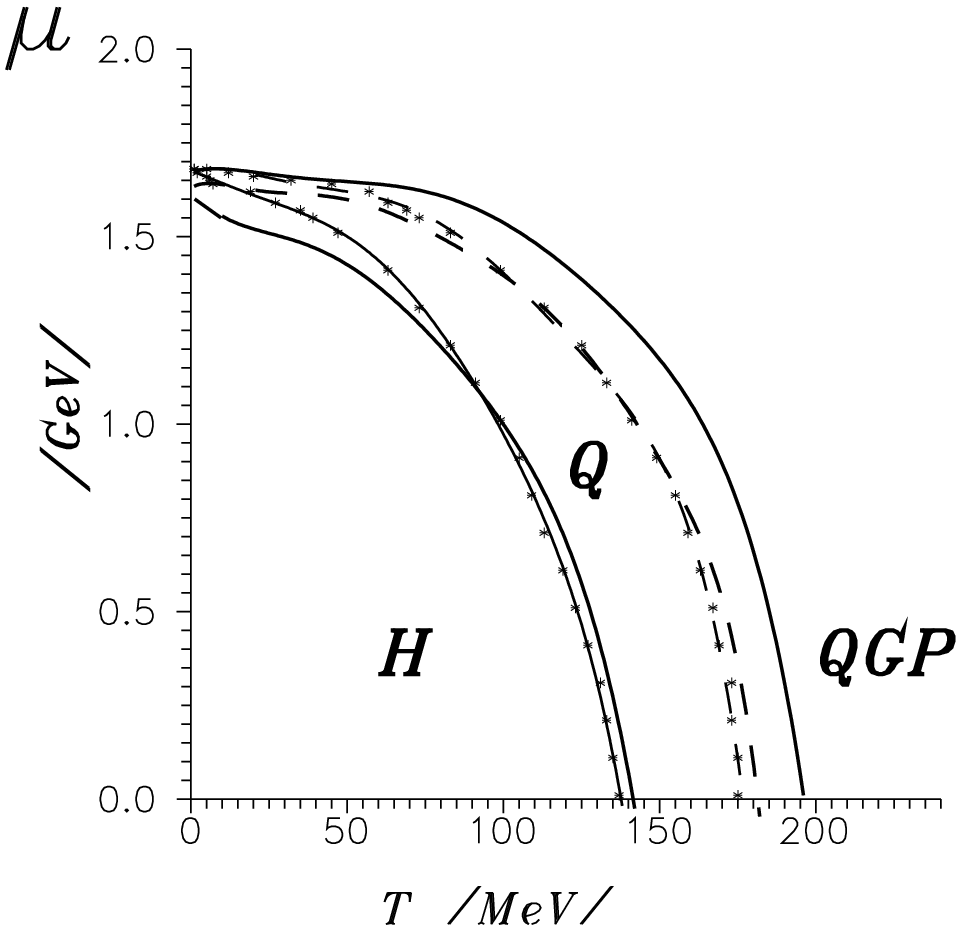}
~\\
~\\
~\\
\begin{center}
{\small Figure 4: DPTM (solid) and SPTM (dashed) transition curves for
HCM with and without (dots) accounting for 30 resonance in the $H$ phase.}
\end{center}
%%%%%%%%%%%%%%%%%%%%%%%%%%%%%%%%%%%%%%%%%%%%%%%%%%
\newpage
%%%%%%%%%%%%%%%%%%%%%%%%%%%%%
\begin{figure}
\hspace*{1.cm}
\input{fig5a.tex}
\end{figure}

\vspace*{1.cm}
%\begin{figure}
\hspace*{.4cm}
\input{fig5b.tex}
%\end{figure}
\begin{center}
{\small
Figure 5: Schematic illustraton for space-time evolution of hot matter
accordind to SPTM (a) and DPTM (b) within Bjorken hydrodynamical version. Initial energy density of the fireball
assumed to be $\epsilon_0 \approx 4$ GeV/fm$^3$.}
\end{center}
%%%%%%%%%%%%%%%%%%%%%%%%%%%%%%%%%%%%%%%5

%%%%%%%%%%%%%%%%%%%%%%%%%%%%%%%%%%%%%%%%%%%%%%%%%%%%%%%%%%%%
\end{document}